# Assessment of the Rankine Vortex Model for Modeling the Gas Entrainment in a Representative Mock-up of a Sodium Fast Reactor

**Guenadou D, Aubert P, and Descamps J-P**
CEA, DES, IRESNE, Nuclear Technology Department, Center of Cadarache
Saint Paul lez Durance, F-13108, France
david.guenadou@cea.fr; philippe.aubert@cea.fr; jean-philippe.descamps@cea.fr

## ABSTRACT

Since decades, the CEA has been involved in the development of 4th generation reactors cooled by sodium. The dedicated experimental platform PLATEAU was erected to study the different issues and built a results database for the code validation. As experiments with sodium are complex, the tests are led in mock-ups using water as a simulant fluid. The MICAS mock-up, model of the upper plenum, aims at studying the gas entrainment at the free surface. Gas is prohibited in the reactor core due to neutron effects, which may induce reactivity variations. Since vortices are difficult to handle by numerical codes in complex geometries such as reactor ones, criteria have been developed based on the Rankine vortex model to forecast the air entrainment. However, in some case, numerical results do not correlate the experimental ones and those discrepancies are may be due to the vortex model. This article aims at comparing the vortices characteristics calculated using the Rankine vortex model with experimental results in order to improve the gas entrainment criteria. The parameters of the Rankine model are determined using the velocity fields measured by the Particle Image Velocity (PIV) method and the vortices diameter is calculated using the raw PIV images.

## 1. INTRODUCTION

At the beginning of the 2000s, the rapid growth of nuclear power worldwide, particularly in China, led to projections indicating a significant rise in uranium costs over the years. Although the cost of raw uranium constitutes only 3 to 6% of the total price of electricity, market tensions could result in shortages and jeopardize electricity production. To ensure France's energy independence, it is crucial to establish ties with uranium-producing states, despite the ore's relatively even distribution across the globe and the fact that its production is not concentrated in a few countries, like oil. In this context, and with a focus on the sustainable management of nuclear materials, France decided in 2006 to develop a reactor prototype that meets the 4th generation criteria defined by the Generation IV International Forum and based on 4 themes: sustainable development, economic profitability, safety and reliability, non-proliferation of nuclear materials. Launched in 2010 under the aegis of the CEA, the ASTRID project aimed at designing a Sodium Fast Reactor (SFR) demonstrator on a sufficiently high scale to extrapolate the developed technologies to an industrial reactor exceeding 1000 MW. Although the ASTRID project was halted in 2019 after the completion of the basic design stage, studies on sodium technology are ongoing at CEA. For the reactor to operate normally, the presence of gas in the core must be avoided, as large volumes of gas entering the fissile part pose a high risk to nuclear safety. An increased void fraction would result in a positive reactivity effect, causing a power peak. Additionally, the heat transfer coefficient would be drastically reduced, potentially leading to inadequate core cooling and fuel deterioration. Furthermore, argon passing through the core's fissile part becomes radioactive due to the strong neutron flux, affecting the accuracy of detecting fission products in the primary fluid linked to potential fuel assembly leaks. Gas presence also harms the primary pumps' operation and can damage them. Unlike the pressurized water reactor, SFRs operate at low pressure (around 10 bar) and have a free surface covered by argon. This has the function of damping variations in the volume of sodium depending on its temperature due to its expansion, the vessel being closed [1]. Gas can be drawn in by the suction of sodium into the Intermediate Heat Exchanger (IHX) [1], [2], [3]. After passing through the primary pumps, it can potentially accumulate in the lower reactor parts, forming a gas pocket. Various gas entrainment modes are possible depending on the free surface's state. Bubbles trapped by the breaking sodium wavelets at the surface can be transported in the flow entering the IHX, leading to a slow gas accumulation. Surface vortices can also be created under certain hydraulic conditions, leading, in the most critical cases, to a column of gas between the free surface and the entrance of the IHX or to the detachment of bubbles from the tip of the vortices. This gas entrainment mode, being the most severe, necessitates studying its characteristics specific to the reactor design.

In terms of gas entrainment issues, especially for SFRs, the primary challenge in numerical simulations lies in modeling the liquid/gas interface in large geometries. To capture small-sized phenomena, such as millimeter-diameter bubbles, the mesh size in the calculation domain must match these dimensions. However, since the calculation domain ranges from meters (in mock-ups) to tens of meters (for reactors), the grid contains a vast number of meshes, significantly impacting computation time. Moreover, two-phase models are complex and prone to misuse, requiring careful analysis and critique of their results. For these reasons, the complete simulation of gas entrainment in a SFR currently remains difficult to envisage. . Some calculations of the free surface in the hot plenum's complete geometry have been undertaken without considering gas entrainment [4]. Therefore, using models to study gas entrainment by vortices seems inevitable.

Due to the complex phenomena involved, forecasting gas entrainment with a general theoretical model is not feasible. The authors of [5] developed a gas entrainment criterion based on a vortex model with experimentally determined parameters. The methodology proposed by [5] suggests to carry out single-phase modeling with a slip boundary condition at the liquid/gas interface and to apply the gas entrainment criterion on the numerically determined velocity fields. This approach significantly reduces computation time, allowing for extensive studies. The parameters of the gas entrainment criterion were determined in an analytical test section [6] that does not accurately represent the actual geometry of a reactor. This methodology shows some discrepancies compared to experimental results obtained from a high-scale mock-up representative of the ASTRID reactor's full geometry [7]. This may be due to the vortex model and this article is dedicated to assess the accuracy of this one based on comparison with experimental results. The first section introduces the vortex and gas entrainment models, the second presents the experimental setup and the methodology and the last part compares the experimental results with the calculations obtained from the vortex model.

## 2. RANKINE VORTEX AND GAS ENTRAINMENT MODELS

As in most cases in fluid mechanics, the characteristics of vortices in terms of velocity cannot be explicitly derived from solving the Navier-Stokes equations. To address this issue, several more or less complex models have been developed using simplifying assumptions. These make it possible to determine the velocity field in the fluid around the vortices, but also to calculate the shape of the liquid/gas interface and therefore the depth of the vortices. These data are useful for the development of gas entrainment models for computer codes. In this section, we will present the common RANKINE vortex model and the gas entrainment criteria developed based on this one.

The model developed by Rankine [8] is based on the hypothesis that the fluid behaves like a rigid body inside a circle of radius $r_0$ with the center of the vortex as its origin ($r_0$ being considered as the radius of the vortex). Outside this circle, the flow is modeled by a potential flow. This approximation is considered valid for sodium flows due to its low viscosity [9]. Equation 1 provides the expression for the velocity fields around the vortex.

$$\begin{cases} v_r = 0 \\ v_\theta = \frac{\Gamma_\infty}{2\pi} r/_{r_0^2} \leftarrow r < r_0 \Big| \; v_\theta = \frac{\Gamma_\infty}{2\pi} 1/r \leftarrow r > r_0 \\ v_z = 0 \end{cases} \tag{1}$$

Where $v_r$, $v_\theta$, $v_z$, are respectively the radial, tangential and vertical velocity, $\Gamma_\infty$ is the circulation.
Using the expression of the velocities in equation 1, the vortex profile can be calculated by integrating the Navier-Stocks equations while assuming constant pressure at the surface. The relationship between the depth of the vortex and its radius from the center is expressed in Equation 2.

$$h(r) = \begin{cases} \frac{\Gamma_\infty^2}{4\pi^2 g r_0^2}\left(1 - \frac{r^2}{2r_0^2}\right), \; r < r_0 \\ \frac{\Gamma_\infty^2}{8\pi^2 g}\frac{1}{r^2} \;, \; r \geq r_0 \end{cases} \tag{2}$$

It is then possible to determine the maximum depth $L_{gc}$ of the vortex according to its parameters as shown in Equation 3.

$$L_{gc} = \frac{\Gamma_\infty^2}{4\pi^2 g r_0^2} \tag{3}$$

A gas entrainment criterion can be determined from the length $L_{gc}$ of the vortex. Gas entrainment is considered to occur if $L_{gc}$ reaches the depth P of the suction nozzle. Numerically, this criterion is met when the $L_{gc}/P$ ratio, as defined in Equation 4, equals 1.

$$\frac{L_{gc}}{P} = \frac{\left(\Gamma_\infty/_{2\pi r_0}\right)^2}{gP} = K.Fr_{\Gamma_\infty}^2 \tag{4}$$

The RANKINE model does not take into account surface tension forces, the criterion developed by of Equation 14 leads to overestimate the depth of the vortices. Indeed, the surface tension acting mainly at the center of the vortex will tend to curve its profile. Since the model overestimates the depth of the vortex, the results are envelope from the safety point of view. However, the influence of the surface tension effect will be nevertheless discussed in the results section.
It should be noted that the gas entrainment criterion in this article is developed with the RANKINE vortex model, whereas the authors of [5] used the BURGER model. However, since the tangential velocities of these two models are similar, the resulting gas entrainment criteria are also comparable.
The criterion, expressed in dimensionless form using local data in Equation 4, depends on a Froude number calculated from the velocity determined from the circulation around the vortex and its radius. The parameter K theoretically equals 1, which means that the number $Fr_{\Gamma_\infty}^2$ must not exceed the value of 1 to prevent gas entrainment. However, this parameter must be adjusted based on experimental results. To achieve this, experiments were conducted using a model representative of the reactor vessel. To assess the vortex model, we require the velocity fields around the vortex and its diameter relative to its depth.

## 3. SIMILITUDE AND EXPERIMENTAL SETUP

Sodium experiments are complex and expensive to carry out due to the high temperatures at which they take place and the danger associated with the violent reaction of sodium with water and air. In addition, the instrumentation and visualization of phenomena are challenging because of sodium's opaque nature. Historically, many experiments have used a simulant fluid, with water being a preferred candidate because it is inexpensive, transparent for measurements, and easy to implement. Part of the studies were carried out on models in water. However, using a simulant fluid requires a similarity approach to ensure the representativeness of the phenomena and the validity of the studies. The dimensionless numbers specific to the reactor and the model must be identical. The main dimensionless numbers characterizing the fluid behavior in the primary circuit [10], [11] are: Reynolds, Weber, Froude, Euler, Richardson,

Peclet. However, in general, it is impossible to keep all these dimensionless numbers simultaneously because their expression differs [12]. For example, the Reynolds number scales linearly with length, while the Froude number is inversely proportional to the square root of length. Therefore, a selection of the relevant dimensionless numbers is necessary. This choice is based on expert considerations or on the analysis of the equations governing the studied phenomenon. Since thermal effects are not considered (the temperature near to the surface is homogenous in the vessel), Peclet and Richardson are not included. The Euler number is also excluded, as the pressure is relatively constant at the surface, it does not affect vortices characteristics. Regarding the number of Reynolds, it has not to be observed if the flow in the model and the reactor are of the same regime (either laminar or turbulent). For free surface studies, the authors of [13] show that the conservation of the numbers of Froude and Weber alone is sufficient for representing the phenomenology. These conditions constrain the size of the model to a 5/8 scale, which is relatively large and leads to expensive and complex experiments. To reduce costs, we maintain the Froude number and allow a distortion in the Weber number by reducing the model size while keeping it at a reasonably large scale, the effects of surface tension being more preponderant for small models [14]. Moreover, as the surface tension mainly affects the center of the vortex, the Weber number primarily influences the vortex shape at this point. Since our studies focus on the vortex edges, close from the free surface, the influence of the Weber number in this area can be considered negligible.
Given this constrains, the MICAS model was designed to maintain geometric similarity with the ASTRID hot plenum. A picture is shown in Figure 1. Its dimensions are approximately 2.5 m in diameter and 1.7 m in height. The 1/6 scale reduction compared to the ASTRID reactor was chosen based on the CEA's experience with this type of model and model and to ensure a sufficiently large size for studying the free surface. This scale also makes it possible to represent the components in detail (such as the Upper Core Structure), although certain geometric simplifications were necessary.
The flow distribution in the ASTRID core is quite complex and replicating this exactly in the MICAS model would be challenging. Therefore, the flow distribution was simplified into 3 zones: fissile zone, reflector, internal storage.

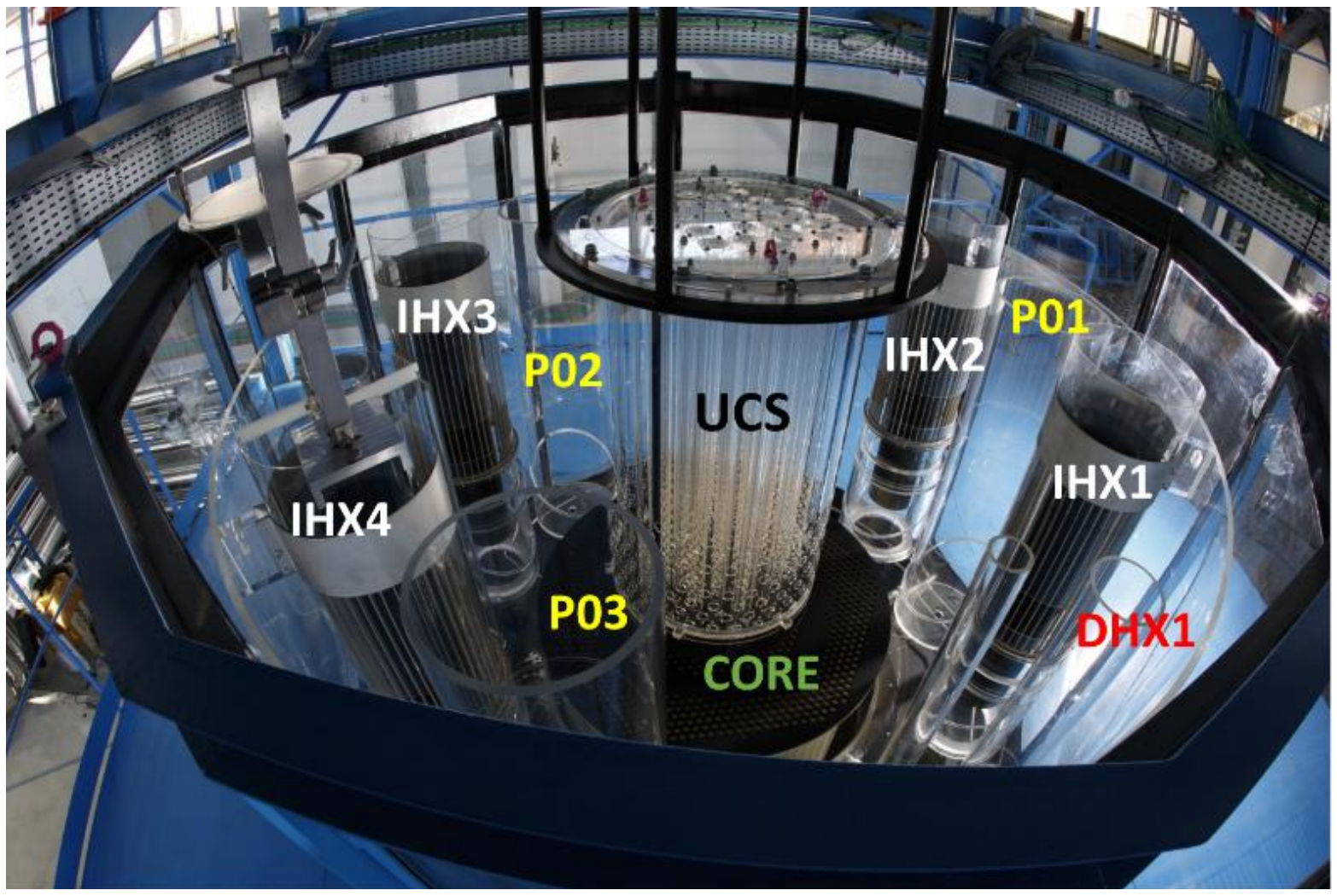

**Figure 1: Top view of the MICAS mock-up (P0x stands for the pump pit, UCS, for the Upper core Structure, IHXx, for the Intermediate Heat Exchanger, DHX1, for the decay heat exchanger)**

The vortices appear naturally at the surface and their shape and the intensity of the gas entrainment depends on the operating parameters (flow, geometry…). Feedback from experiments carried out at the CEA has shown that vortices only occur in certain risk zones where vorticity and downward velocity exist together. To ensure the MICAS model accurately represents the reactor's flow, the flow rate should be calculated using the Froude number, maintaining its equivalence between the reactor and the model.

However, since this study aims to assess the RANKINE vortex model, it only requires local characteristics of the flow, such as the velocity field at the surface and vortex features (e.g., diameter). These characteristics can be obtained under various operating conditions, as long as vortices with different behaviors (with and without gas entrainment) occur at the surface. However, to establish a starting point for the studies, the initial operating conditions of the model during the experimental campaign were calculated by maintaining the Froude number. From this baseline, the flow rate was adjusted to produce different types of vortices. Nevertheless, these adjustments were kept within a range close to the operating conditions calculated using similarity to preserve a minimum level of representativeness.
The flow velocity in the MICAS model is measured using the Particle Image Velocity (PIV) technique, which allows large 2D velocity fields to be acquired. This technique is already implemented worldwide to measure the velocity in reactor models [15]. It is based on tracking the movement of particles illuminated by two short laser pulses separated by a brief delay. An Evergreen™ double pulse laser (wavelength, duration and energy, frequency: 532 nm, 5 ns, 200 mJ, 15 Hz) is mounted to a 3-axis movement table to achieve positioning accuracy of 1 mm. It generates a thin horizontal sheet of light (1 to 2 mm thick) 2 cm below the free surface. The MICAS model uses clean deionized water seeded with 1 µm nylon particles, which have the same density as water and follow the flow accurately. Images are recorded using a PowerView™ Plus Charge Coupled Device (CCD) camera with spatial resolution 2048 × 2048 pixels. The camera is located perpendicular to the laser sheet, above the free surface. The occurrence of vortices at the surface is stochastic, making it impossible to predict their exact location. However, for given operating conditions, they consistently appear in the same area. This consistency allows for strategic placement of the camera. The PIV recording is triggered when a vortex enters the measurement domain. A bandpass filter at the laser wavelength is installed on the camera to capture only the light scattered by the particles, thereby reducing noise from ambient light. Data were analyzed using Insight 4G™ software (version 10.10.4). Spatial calibrations, which allow pixels to be converted into millimeters, are carried out prior to measurement using a target. The measurement domain is divided into a grid defining the interrogation windows for the cross-correlation calculations. The velocity vector file is calculated using two-pass processing. For the first pass, the size of the interrogation window is 128x128 pixels and in the second pass, it is reduced to 64x64 pixels. This approach speeds up the calculation and provide more points in the velocity field. In each experiment, we ensure that the interrogation windows contain a minimum of 9 to 10 particles to obtain quality results. The delay between the two laser pulses varies from 4 to 5 ms and is adjusted according to the flow velocity conditions so that the maximum displacement does not exceed 1/3 of the smallest window size of the interrogation region, or 10 pixels in our case. The final instantaneous velocity fields are smoothed using a moving average over a 3 × 3 cell box. The Insight 4G software™ allows to calculate the uncertainties of the instantaneous velocity using the Charonko & Vlachos method [16] which leads to about 2% of error on the velocity magnitude in the experimental conditions of this study. This error is valid as long the PIV guidelines are followed. In this type of experiment, conducting reproducibility tests is not strictly possible because identical operating conditions do not consistently produce the exact same vortices, as their occurrence at the surface is stochastic.

## 4. ACCURACY OF THE VORTEX MODEL VERSUS EXPERIMENTAL RESULTS

The left image in Figure 2 shows a typical case where bubbles are seen detaching from the vortex tip. As mentioned in the experimental setup, the laser sheet is positioned at 20 mm below the free surface. This value was chosen with regard to studies by [17], [18] where the velocity was measured between 10 and 45 mm below the liquid/gas interface. Since the camera is located above and perpendicular to the water surface, we assumed that, the slightly undulating behavior of the surface does not alter the recorded images and has no impact on the velocity accuracy. The measurement range is quite wide (140 mm x 140 mm) thanks to the 35mm lens set up on the camera. The right picture of Figure 2 is a raw image of a PIV measurement where we can recognize the scatted light from the seeded particles. For each measurement, we ensure there are enough particles and that their displacement falls within the correct range to obtain accurate velocity results. In the right picture of Figure 2, as the particles are only contained in water, the air core of the vortex can be identified. Using the mm/pixel calibration, this allows measuring its diameter simultaneously with the velocity around it. The experimental results show that the vortex diameter at 20 mm below the surface is between 12 and 22 mm. However, the velocity measurement domain is reduced to the area between the laser source and the vortex. On the right picture of the Figure 2, the region behind vortex core shows a black area due the change of the refraction index from water to air where no particles can be identified for the PIV technic.

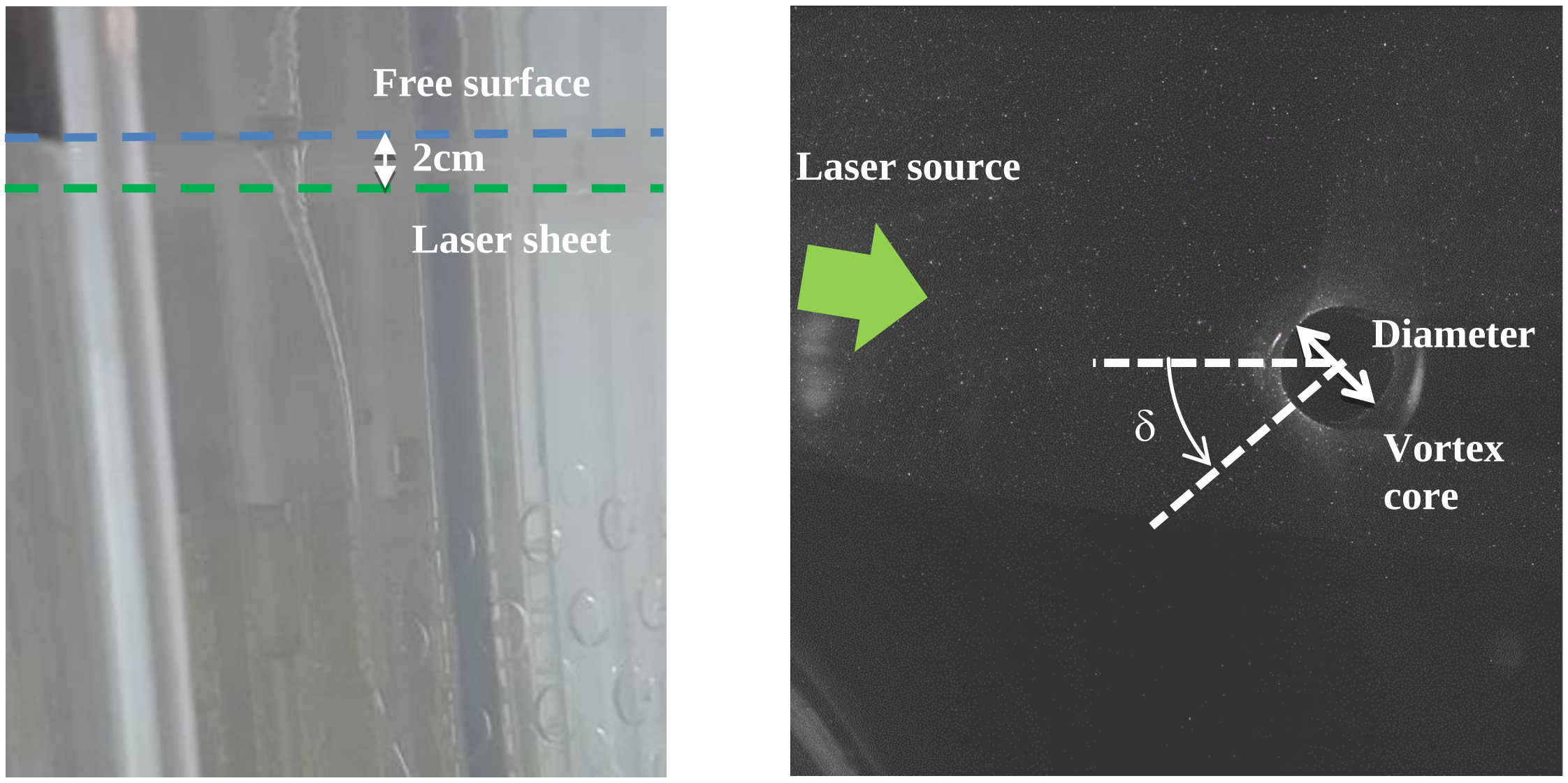

**Figure 2: Front (left picture) and top (right picture) views of a vortex at the free surface.**

The graphs in Figure 3 display a bundle of green curves representing the results at different angles δ (see Figure 2) around the center of the vortex. In all our analyses, these velocities were plotted over the range 0-360° with a step of 5° excluding areas behind the vortex. The blue curves indicate the average across all the angles studied, i.e. the average of the green curves. Notably, we notice disparities in both velocity components considered. The radial velocity is very small compared and close to 0 which is the expected value of the Rankine model. The average of the tangential component shows little dispersion and is relatively smooth, while that of the radial velocity exhibits more fluctuations. In Figure 3, the experimental tangential velocity decreases to a minimal value from the vortex center and increases beyond a specific radius. However, as the tangential velocity of the Rankine model in Equation 1 reaches an absolute maximal value at $r_0$, it is possible to determine this parameter from the PIV measurements. The circulation ($\Gamma_\infty$) is calculated from Equation 1 and using the value of the measured velocity at the radius $r_0$. Another method to determine the Rankine parameter involves fitting the mean curve using a least square method. The red curves (dashed and solid lines) on the tangential velocity graph are plotted based on the Rankine model in Equation 1, with parameters determined by both methods. The first method, based on the absolute maximal tangential component, tends to underestimate the velocity (in absolute values), while the second method overestimates the values at $r_0$. Using these Rankine parameters, it is possible to plot the theoretical depth of the vortex in relation to its radius and compare it with the experimental diameter measured from the raw PIV image.

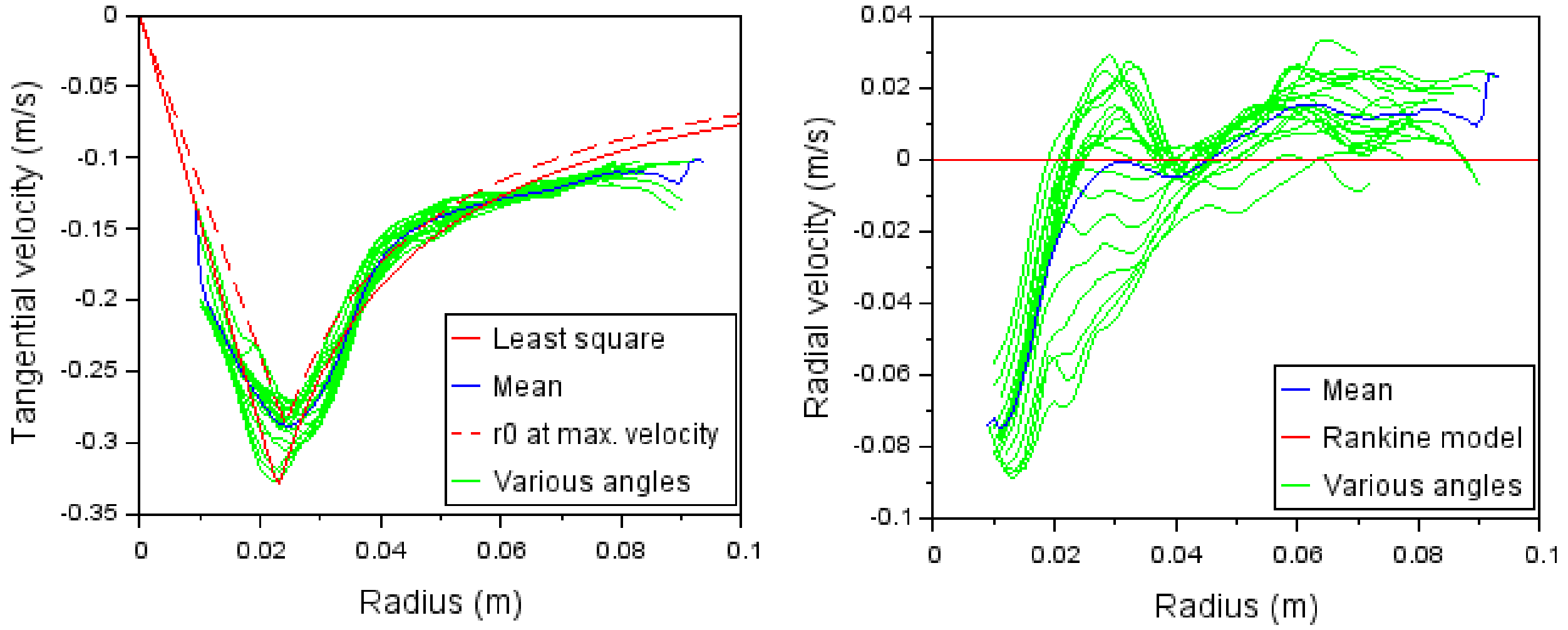

**Figure 3: Tangential and radial velocities from the vortex center - at various angles (green curves), mean value (blue curve), Rankine model parameters calculated from least square method, Rankine model parameters determined assuming r0 at maximal absolute velocity.**

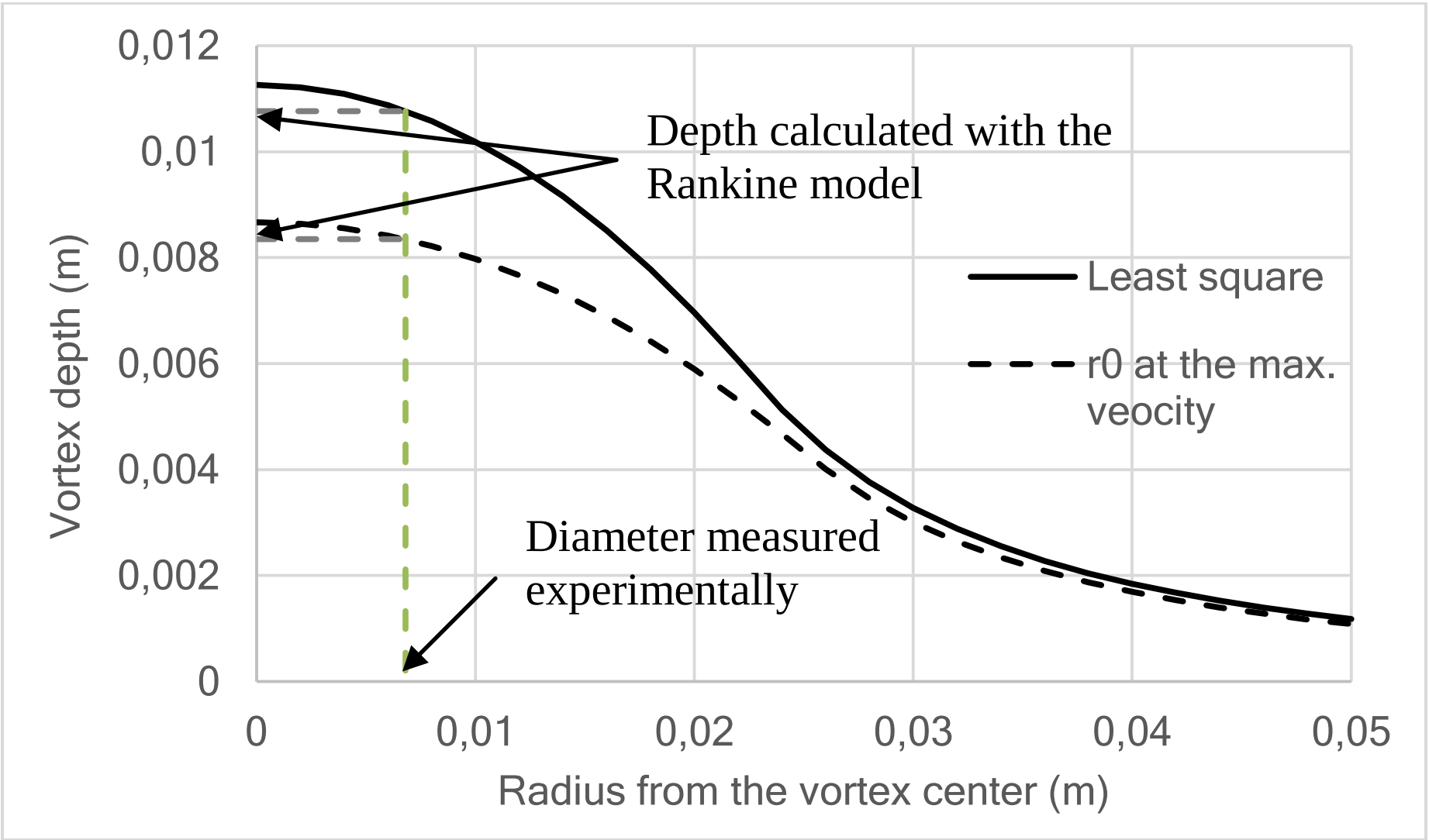

**Figure 4: Vortex depth versus the radius from vortex center calculated with the Rankine model**

The Figure 4 illustrates the vortex profile determined using the Rankine model, with parameters calculated from the velocity measurement obtained PIV in Figure 3. It is evident from Figure 4 that the curve does not reach 20 mm at any the radius, indicating that the Rankine model underestimates the depth of the vortex. This result is consistent across all measurements from the experimental campaign, as shown on Figure 5. The vortex depths in the measurement plane range from approximately 5 mm to 15 mm, while their expected size is 20 mm. These calculations account for the variations in vortex properties around the vortex itself (the δ angle in Figure 2) and are determined using the green bundle curves shown in Figure 3. There are two possible explanations for this discrepancy: either the Rankine model cannot accurately simulate the vortex shape in the studied geometric, or the experimental conditions differ slightly from those used to calculate the depth. From the point of view of the first hypothesis, the Rankine model is known to accurately simulate flows in very simple geometry such as cylinders. However, the IHX outlet geometry of the MICAS mock-up may involve complex phenomena that are not captured by this vortex model. In the other point of view, slight shifts in the elevation or angle of the laser sheet relative to the horizontal could affect the measurement location. Additionally, the sensor used to regulate the water depth (guided radar technology) has an error margin of ±2 mm. Analyzing the vortex depth calculated using the experimentally measured diameter, as shown in Figure 5, reveals that the results are about 10 mm below the expected value (20 mm, the experimental depth of the laser sheet). This suggests that the measurement method may need to be improved to enhance the accuracy of the overall measurement chain.

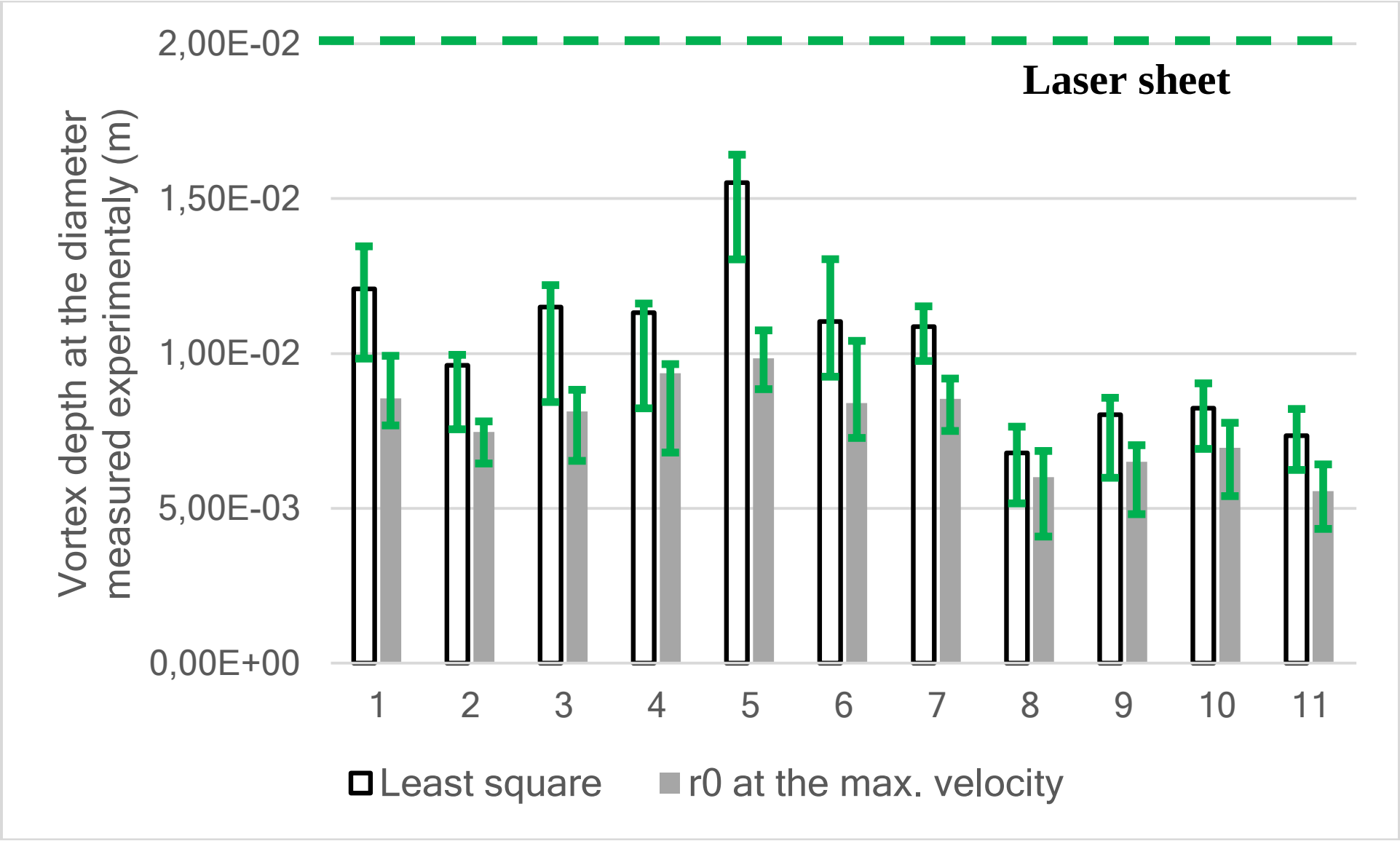

**Figure 5: Vortex depth calculated with the Rankine model using the experimental measurements of the diameter at 20 mm below the surface. The green vertical bars represent the range of the vortices depth calculated using the green bundle curves of the figure 3.**

In addition, these results allow us to analyze the effect of surface tension on the shape of the vortices. While this analysis may not improve our model, as surface tension tends to reduce vortex depth, it is still valuable to discuss its effects from a general perspective. Equation 2 can be rewritten to incorporate the surface tension force, as defined in Equation 5, which depends on the variation of the vortex curvature.

$$f_\sigma = -\sigma \frac{d\kappa}{dr} \tag{5}$$

Where σ is the surface tension, r the radius of the vortex and κ the curvature, which can be calculated from the equation of the shape of the vortex.
By setting R=r/r0 and H=h/P, Equation 2, including the surface tension, becomes Equation 6. In this equation, the curvature is negative, due to the chosen coordinate system.

$$H(R) = \begin{cases} Fr_{\Gamma\infty}^2 \left(1 - \frac{R^2}{2}\right) + \frac{Fr_{\Gamma\infty}^2}{We_{\Gamma\infty}} \mathrm{K}(R), |R < 1 \\ Fr_{\Gamma\infty}^2 \cdot \frac{1}{2R^2} + \frac{Fr_{\Gamma\infty}^2}{We_{\Gamma\infty}} K(R), |R \geq 1 \end{cases} \tag{6}$$

In our study, for all the vortices analyzed, the radius of the vortices measured using the raw PIV images is smaller than the radius calculated using the velocity measurement. This indicates that, in our measurement location, the shape of the vortex are determined by the upper formula (R<1) of Equation 6. This equation is divided into 2 terms: the first represents the effect of the inertia and gravity, while the second accounts for the tension surface force. By comparing the magnitude of these terms, we can determine whether surface tension should be considered when calculating the vortex depth in our specific case. However, since we do not have the exact equation for the curvature, in first approach, we approximate it using the vortex depth defined in Equation 2. Our calculations show that the second term, specific to surface tension, is approximately 20 times smaller than the term driven by inertia and gravity. Therefore, it appears that in this study, surface tension does not significantly affect the shape of the vortices.

## 5. CONCLUSION

The 1/6th scaled MICAS mock-up was designed to study gas entrainment behavior in the upper plenum of a SFR reactor. To achieve similar surface flow behaviors, the reactor Froude number was replicated in the MICAS mock-up by adjusting the experimental operating conditions. Gas entrainment primarily results from vortices occurring at the free surface. Numerical codes struggle to simulate these vortices in large geometries due to complex phenomena and lengthy computation times. Therefore, a criterion based on the Rankine model has been developed and implemented in the codes to determine the onset of gas entrainment. This criterion assumes that gas entrainment occurs when the vortex tip reaches the IHX inlet. However, this methodology shows some discrepancies when compared with experimental results obtained from the MICAS mock-up. One hypothesis to explain these discrepancies is the potential inadequacy of the Rankine model for accurately simulating the flow.
To assess the accuracy of the Rankine model, vortices characteristics in terms of diameter and velocity were measured using the PIV technic in a plane at 20 mm below the free surface. The results indicate that the Rankine model underestimates the depth of the vortices. However, uncertainties regarding the exact location of the laser sheet could explain the lack of correlation between the model and the experimental results. New measurements are planned in simpler geometries to better control the experimental parameters and verify this analysis. Furthermore, velocity fields and vortex diameters will be measured simultaneously in two horizontal planes at different depths to further validate or refute the Rankine model. Additional discussions focus on the effects of surface tension. An analysis of the system's equations shows that surface tension can be neglected under the specific conditions of our study.